\documentclass[sigconf]{aamas} 
\usepackage{balance} 
\usepackage[ruled,linesnumbered]{algorithm2e}
\usepackage{subcaption}
\usepackage{latexsym}
\hypersetup{
    colorlinks=true,
    linkcolor=blue,
    filecolor=magenta,      
    urlcolor=blue,
    citecolor=blue,
}
\usepackage{multirow}
\usepackage{url}
\usepackage{stfloats}

\setcopyright{ifaamas}
\acmConference[AAMAS '22]{preprint}{}
{}{}
\copyrightyear{2022}
\acmYear{2022}
\acmDOI{}
\acmPrice{}
\acmISBN{}

\acmSubmissionID{6}

\title[AAMAS-2022 Formatting Instructions]{Policy Regularization via Noisy Advantage Values for Cooperative Multi-agent Actor-Critic methods}

\author{Jian Hu$^{*+}$}
\affiliation{
  \institution{National Taiwan University}
  \city{Taipei}
  \country{Taiwan}}
\email{janhu9527@gmail.com}

\author{Siyue Hu$^{*}$}
\affiliation{
  \institution{National Taiwan University}
  \city{Taipei}
  \country{Taiwan}}
\email{husiyuehusiyue@gmail.com}

\author{Shih-wei Liao}
\affiliation{
  \institution{National Taiwan University}
  \city{Taipei}
  \country{Taiwan}}
\email{liao24@gmail.com}

\thanks{$^*$ Jian Hu and Siyue Hu contributed equally to this work.}
\thanks{$^+$ Corresponding Author.}

\begin{abstract}
Recent works have applied the Proximal Policy Optimization (PPO) to the multi-agent cooperative tasks, such as Independent PPO (IPPO); and vanilla Multi-agent PPO (MAPPO) which has a centralized value function. However, previous literature shows that MAPPO may not perform as well as Independent PPO (IPPO) and the Fine-tuned QMIX on Starcraft Multi-Agent Challenge (SMAC). MAPPO-Feature-Pruned (MAPPO-FP) improves the performance of MAPPO by the carefully designed agent-specific features, which may be not friendly to algorithmic utility. By contrast, we find that MAPPO may face the problem of \textit{The Policies Overfitting in Multi-agent Cooperation(POMAC)}, as they learn policies by the sampled advantage values. Then POMAC may lead to updating the multi-agent policies in a suboptimal direction and prevent the agents from exploring better trajectories.
In this paper, to mitigate the multi-agent policies overfitting, we propose a novel policy regularization method, which disturbs the advantage values via random Gaussian noise. The experimental results show that our method outperforms the Fine-tuned QMIX, MAPPO-FP, and achieves SOTA on SMAC without agent-specific features. We open-source the code at \url{https://github.com/hijkzzz/noisy-mappo}.
\end{abstract}


\newcommand{\BibTeX}{\rm B\kern-.05em{\sc i\kern-.025em b}\kern-.08em\TeX}

\begin{document}


\pagestyle{fancy}
\fancyhead{}


\maketitle 


\section{Introduction}
Multi-Agent Reinforcement Learning (MARL) has seen revolutionary breakthroughs with its successful application to multi-agent cooperative tasks such as robot swarms control \citep{huttenrauch2017guided}, autonomous vehicle coordination~\citep{cao2012overview} and computer games ~\citep{samvelyan2019starcraft}. As for scalability and communication security problems, decentralized execution of multi-agent policies that act only on their local observations is widely used. An intuitive approach for decentralized multi-agent policy learning is the Independent Q Learning (IQL) ~\citep{tan1993multi}. However, IQL does not address the non-stationarity introduced due to the changing policies of the learning agents. Thus, unlike single-agent Reinforcement Learning (RL) algorithms, there is no guarantee of convergence even at the limit of infinite exploration. Therefore, the \textit{Centralized Training and Decentralized Execution (CTDE)} \citep{kraemer2016multiagent}, which allows for agent to access global information during training stage, is widely used in MARL algorithms \citep{lowe2020multiagent, rashid2018qmix}.

Many CTDE algorithms, e.g. MADDPG \citep{lowe2020multiagent}, MAAC \citep{iqbal2019actorattentioncritic}, QMIX \citep{rashid2018qmix} , have been proposed for multi-agent cooperative tasks. Among these algorithms, the finetuned QMIX ~\citep{hu2021riit} achieves the SOTA performance in the popular MARL benchmark environment Starcraft Multi-Agent Challenge(SMAC)~\citep{samvelyan2019starcraft}. To enable effective CTDE for multi-agent Q-learning, the Individual-Global-Max (IGM) principle~\citep{son2019qtran} of equivalence of joint greedy action and individual greedy actions is critical.
QMIX ensures that the IGM condition holds by the mixing network with \textit{Monotonicity Constraint} \citep{rashid2018qmix}. However, the mixing network leads to limitations in its scalability, and monotonicity constraints prevent it from learning correctly in non-monotonic environments ~\citep{son2019qtran}. We turn our attention to the efficient single-agent RL algorithms, such as Trust Region Policy Optimization (TRPO) \citep{schulman2015trust} and Proximal Policy Optimization (PPO) ~\citep{schulman2017proximal}, as their unlimited expressive power and high sample efficiency. 

Recently, the literature \citep{de2020independent} applies the PPO to the multi-agent tasks directly, called Independent PPO (IPPO);  literature \citep{de2020independent} also extend IPPO to the vanilla Multi-agent PPO (MAPPO) which using the centralized value function. However, the MAPPO  may not perform as well as Independent PPO (IPPO) and the Fine-tuned QMIX \citep{hu2021riit}. Then literature \citep{yu2021surprising} further improves the performance of MAPPO and IPPO by carefully designed agent-specific features, which may be not friendly to algorithmic utility, called MAPPO-Feature-Pruned (MAPPO-FP) (details in Sec. \ref{sec:agent_specific}); literature \cite{li2020multi} proposed a multi-agent TRPO algorithm, but only for the case where each agent has a private reward. By contrast, we find that MAPPO may face the problem of \textit{The Policies Overfitting in Multi-agent Cooperation(POMAC)} as they learn policies by the sampled centralized advantage values \citep{mnih2016asynchronous}. Then POMAC may lead to updating the policies of some agents in a suboptimal direction and prevent the agents from exploring better trajectories. 

In this paper, (1) To mitigate multi-agent policies overfitting, we propose two policy regularization methods, i.e, Noisy-Value MAPPO (NV-MAPPO) and Noisy-Advantage MAPPO (NA-MAPPO), which disturb the advantage values via random Gaussian noise. (2) Empirical results show that our approaches achieve better performance than the Fine-tuned QMIX \citep{hu2021riit}, MAPPO-FP \citep{yu2021surprising}, and is much better than MAPPO, achieving SOTA performance in SMAC. 

Although our method is very simple, it shows that noise perturbation of the advantage function can significantly improve the performance of the multi-agent actor-critic algorithms. Interestingly, our findings demonstrate that replacing agent-specific features (in MAPPO-FP) with noise may make MAPPO works better.
\section{Background} \label{sec:background}

\textbf{Dec-POMDP} We consider a cooperative task, which can be described as a decentralized partially observable Markov decision process (Dec-POMDP)\citep{ong2009pomdps}. The cooperative agent chooses sequential actions under partial observation and environment stochasticity.
Dec-POMDP is a tuple $(\mathcal{S},\mathcal{A},\mathcal{O},\mathcal{R},\mathcal{P},n,\gamma$) where $\mathcal{S}$ is state space. $\mathcal{A}$ is joint action space. $o_i = \mathcal{O}(s;i)$ is partially observation for agent $i$ at global state $s$. $\mathcal{P}(s'|s,\mathcal{A})$ is the state transition probability in the environment given the joint action $\mathcal{A} = (a_1, \ldots, a_N)$. Every agent has same shared reward function $\mathcal{R}(s,\mathcal{A})$. $N$ denotes the number of agents and $\gamma \in [0,1)$ is the discount factor. The team of agents attempt to learn a joint policy  $\boldsymbol{\pi} = \langle \pi_{1},... ,\pi_{N}\rangle$ that maximises their expected discounted return.
\begin{eqnarray}\label{jointobject}
V^{\pi}\left(s_0\right)=\mathbb{E}_{a^{1} \sim \pi^{1}, \ldots, a^{N} \sim \pi^{N}, s \sim T}\left[\sum_{t=0}^{\infty} \gamma_{t} r_{t}\left(s_{t}, a_{t}^{1}, \ldots, a_{t}^{N}\right)\right]
\end{eqnarray}

\textbf{CTDE}
Centralized training with decentralized execution(CTDE) paradigm\citep{kraemer2016multiagent}, in which agents can obtain additional information and centralized joint learning; while in the testing phase, agents make the decision based on their own partially observation. Next, we introduce some CTDE algorithms for the multi-agent credit assignment \citep{chang2004all}.

\textbf{Credit assignment} Multi-agent credit assignment \citep{chang2004all} is a critical challenge:in cooperative settings, joint actions typically generate only global rewards, making it difficult for each agent to deduce its own contribution to the team's success. Many CTDE algorithms have been proposed to solve this problem: COMA \citep{foerster2017counterfactual} trains decentralized agents by a centralized critic with counterfactual advantages. MADDPG \citep{lowe2020multiagent} and MAAC \citep{iqbal2019actorattentioncritic} trains a joint critic to extend DDPG \citep{lillicrap2015continuous} to the multi-agent setting, which can be seen as implicit credit assignment \citep{zhou2020learning}. VDN \citep{sunehag2017valuedecomposition}, QMIX \citep{rashid2018qmix} (details in Sec. \ref{sec:related}) decompose the joint action-value function $Q_{tot}$ to individual action-value functions $Q_{i}$ by the value mixing networks. However, the monotonicity constraints limit the expressive power of QMIX, which may learn error argmax action in nonmonotonic cases \cite{son2019qtran} \cite{mahajan2020maven}. Besides, we consider a task including millions of agents, but only several states, the mixing network faces the problem of explosion in the size of $Q_i$. 

\textbf{Policy Gradient (PG)} Then, we briefly introduce the Policy Gradient (PG) and Proximal Policy Optimization (PPO) in single-agent RL. In the on-policy case, the gradient of the object value function $V^{\pi}(s_0) \stackrel{\text { def }}{=} \mathbb{E}_{\pi}\left[\sum_{t \geq 0} \gamma^{t} r_{t}\right]$, where $\gamma \in[0,1)$ with respect to some parameter of the policy $\pi$ is

\begin{equation}
\nabla V^{\pi}\left(s_{0}\right)=\mathbb{E}_{\pi}\left[\sum_{t \geq 0} \gamma^{t} \nabla \log \pi\left(a_{t} \mid s_{t}\right) A^{\pi}\left(s_{t}, a_{t}\right)\right]
\label{eqn:onpolicy_pg}
\end{equation}

where $
A^{\pi}\left(s_{t}, a_{t}\right):= Q^{\pi}\left(s_{t}, a_{t}\right) - V^{\pi}\left(s_{t}\right)
$ is the advantage value function \citep{mnih2016asynchronous} of policy $\pi$, where $Q^{\pi}\left(s_{t}, a_{t}\right) := r_t + \gamma V^{\pi}\left(s_{t+1}\right)$ is the state action value function. Intuitively, PG makes the policy $\pi$ closer to the actions with large advantage value by gradient ascending.

\textbf{Proximal Policy Optimization (PPO)} To improve the sample efficiency of PG, Trust Region Policy Optimization (TRPO) \citep{schulman2015trust} aims to maximize the objective function $V^{\pi}(s_0)$ subject to, trust region constraint which enforces the distance between old and new policies measured by KL-divergence to be small enough, within a parameter $\delta$,


\begin{equation}
\begin{aligned}
J^{TRPO} = \mathbb{E}_{a_{t} , s_{t} \sim \pi_{old}}\left[\frac{\pi\left(a_{t} \mid s_{t}\right)}{\pi_{old}\left(a_{t} \mid s_{t}\right)} A^{\pi_{old}}\left(s_{t}, a_{t}\right) \right]
\end{aligned}
\end{equation}

where $\frac{\pi\left(a_{t} \mid s_{t}\right)}{\pi_{old}\left(a_{t} \mid s_{t}\right)}$ is the Importance sampling (IS) weight, with KL-divergence constraint,

\begin{equation}
\mathbb{E}_{s \sim \rho^{\pi_{old}}} \left[D_{\mathrm{KL}}\left(\pi_{\theta_{\mathrm{old}}}(. \mid s) \| \pi_{\theta}(. \mid s)\right] \leq \delta\right.
\label{eqn:kl_divergence}
\end{equation}

where $\rho^{\pi_{old}}$ is the discounted state distribution \citep{schulman2015trust} sampled by policy $\pi_{old}$. \citep{schulman2015trust} prove that $J^{TRPO}$ is equivalent to the Natural Policy Gradient (NPG) \citep{kakade2001natural}, which enable the the gradient in the  steepest direction of object function. However, in large-scale neural networks, the KL-divergence constraint causes the objective function to be difficult to solve. Therefore, PPO-clip \citep{schulman2017proximal}  proposes an approximate objective function (Eq. \ref{eqn:ppo}),

\begin{equation}
r=\frac{\pi(a \mid s)}{\pi_{{\text {old }}}(a \mid s)}
\end{equation}

\begin{eqnarray}
\begin{aligned}
J^{PPO-clip} & = \\
& \mathbb{E}_{a_t, s_t \sim \pi_{old}}\left[\min \left(r A^{{\mathrm{old}}}(s, a),  \operatorname{clip}(r, 1-\epsilon, 1+\epsilon) A^{{\mathrm{old}}}(s, a)\right)\right]
\end{aligned}
\label{eqn:ppo}
\end{eqnarray}

The function $\operatorname{clip}(r, 1-\epsilon, 1+\epsilon)$ clips the ratio to be no more than $1-\epsilon$ and no less than $1+\epsilon$, which approximates the KL-divergence constraint.
\section{Related Works} \label{sec:related}
In this section we briefly introduce some related work, such as Independent PPO (IPPO), MAPPO-Feature-Pruned \cite{yu2021surprising} and multi-agent TRPO \cite{li2020multi}.

\textbf{Indepent PPO (IPPO) \& Non-stationarity} IPPO train an independent PPO agent for each agent in the multi-agent system, and the literature \cite{de2020independent} shows that he works effectively as well in some multi-agent tasks. However, applying the single-agent policy gradient algorithms to the multi-agent faces the problem of environmental \textbf{non-stationarity}. Specifically, for a certain agent $i$ in a multi-agent system, we can treat other agents' policies as part of the environment; then, the Bellman Equation is
\begin{equation}
 V^{\pi^i}\left(s\right) = \sum_{a} \pi^i(a^i \mid s) \sum_{s^{\prime}, r} p\left(s^{\prime}, r \mid s, a^i, \vec{\pi^-}\right)\left(r+v_{\pi^i}\left(s^{\prime}\right)\right)
\end{equation}

where $p\left(s^{\prime}, r \mid s, a, \vec{\pi^-}\right)$ is the state transition function in the multi-agent setting, and $\vec{\pi^-}$ denotes the policies of other agents. Since the policy of each agent is updated synchronously, the state transition function $p$ is non-stationary, and thus the convergence of the Bellman Equation cannot be guaranteed.

\textbf{MAPPO} extends IPPO's independent critics to a centralized value function with global information $s$. As for the global information, the centralized critic is more accurate than the independent critics. However, literature \cite{de2020independent} demonstrates that it does not work well in some complex environments.

\begin{figure}[htbp]
  \centering
  \includegraphics[width=0.8\columnwidth]{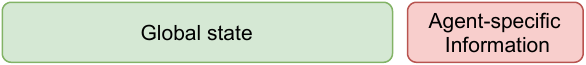}
  \caption{\centering{MAPPO-FP concatenates artificially designed agent's information with global state.}} 
  \label{fig:mappo-fp}
\end{figure}

\textbf{MAPPO-Feature-Pruned (MAPPO-FP)} \label{sec:agent_specific} \cite{yu2021surprising} finetunes the hyperparameters of MAPPO to enable it to perform well in complex multi-agent tasks such as SMAC. \textbf{MAPPO-FP feeds well-designed artificial features (agent-specific features) to the critic networks}, which significantly improved MAPPO's performance in SMAC. The agent-specific features (shown in Figure \ref{fig:mappo-fp}) concatenate the global state $s$ with agent-specific information, such as agent actions mask and agent's information. 

\textbf{multi-agent TRPO} Recently, \cite{li2020multi} propose a 
multi-agent TRPO algorithm with a theoretical analysis. However, this algorithm can only optimize decentralized policies based on local observations and \textbf{private rewards} for each agent, which may not be suitable for complex cooperative tasks with shared rewards. Our method only needs the shared reward without credit assignment.
\section{Preliminaries}
In this section, we first analyze the objective functions of MAPG and MAPPO and then pose the problem of POMAC.

\textbf{Multi-agent PG (MAPG)}
MAPG trains the policies of $N$ agents with a shared advantage function, as shown in Eq. \ref{eqn:multi_pg},

\begin{equation}
\begin{aligned}
    g&= \sum_i^N  \mathbb{E}_{a_{t} , s_{t} \sim \pi}\left[ \nabla \log \pi^i\left(a_{t}^i \mid \tau_{t}^i\right) A( s_t,\vec{a_t})\right]
\end{aligned}
\label{eqn:multi_pg}
\end{equation}

where the shared advantage values are estimated with a centralized value function,

\begin{equation}
A^{\pi}\left(s_{t}, a_{t}\right) = r_t + \gamma V^{\pi}\left(s_{t+1}\right) -V^{\pi}\left(s_{t}\right)
\end{equation}

Fortunately, CTDE allows us to train a central value function using global information $s$. 

\textbf{Multi-agent PPO (MAPPO)} 
We consider a multi-agent TRPO objective function,

\begin{align}
J^{TRPO} = \frac{1}{N} \sum_i^{N} \mathbb{E}_{\vec{a_{t}} , s_{t} \sim \pi_{\text {old }}}\left[ \frac{ \pi^i\left(a_{t}^i \mid \tau_{t}^i\right)}{\pi^i_{{\text {old }}}\left(a_{t}^i \mid \tau_{t}^i\right)} A^{old}(s_t, \vec{a_t})\right]
\end{align}

\begin{equation}
\mathbb{E}_{s \sim \rho^{\pi_{old}}} \left[ D_{\mathrm{KL}}\left(\pi^i_{\theta_{\mathrm{old}}}(. \mid \tau^i) \| \pi^i(. \mid \tau^i)\right] \leq \delta\right.
\label{eqn:kl_sum}
\end{equation}

Then, we can use PPO-clip to optimize the independent policies of the agents, called \textbf{vanilla Multi-agent PPO (MAPPO)},

\begin{equation}
r^i=\frac{\pi^i_{}(a^i \mid \tau^i)}{\pi^i_{{\text {old }}}(a^i \mid \tau^i)}
\end{equation}

\begin{align}
J^{MAPPO-clip} &= \notag\\
& \frac{1}{N} \sum_{i}^N \mathbb{E}_{\vec{a_{t}} , s_{t} \sim \pi_{\text {old }}}\left[ \min \left(r^i A^{{\mathrm{old}}},  \operatorname{clip}(r^i, 1-\epsilon, 1+\epsilon) A^{{\mathrm{old}}}\right)\right]
\label{eqn:ppo2}
\end{align}

Since there is no monotonicity constraint in the actor-critic methods, the expressiveness of MAPPO and MAPG is not limited. 

\textbf{POMAC} Then we can obtain the expected policy gradient of MAPG and MAPPO for an agent $i \in N$,

\begin{equation}
\begin{aligned}
    \frac{\partial \hat{J}}{\partial  \pi^i\left(a_{t}^i \mid s_{t}\right)} &\propto \mathbb{E}_{\vec{a^{j \ne i}} \sim \pi}[ A^{\pi}\left(s_{t}, a^i_t, \vec{a^{j \ne i}}\right)] \\
     &= \mathbb{E}_{\vec{a^{j \ne i}} \sim \pi}[
     r(s_t, a^i_t, \vec{a^{j \ne i}}) + V(s_{t+1}) - V(s_t)]
\end{aligned}
\label{eqn:marginal}
\end{equation}

Interestingly, as this gradient expects the actions of agents other than agent $i$, its value can represent the contribution of agent $i$. Thus this marginal advantage function may be seen as a \textbf{implicit multi-agent credit} \citep{foerster2017counterfactual, sunehag2017valuedecomposition, rashid2018qmix} for agent $i$. 

However, in practice, MAPG and MAPPO estimate this gradient via sampling. According to the large number law, we need a large number of samples with the state $s$ to estimate this expected gradient accurately. When the number of agents $N$ is large, it is almost impossible for us to traverse the action space of all the agents in a batch of samples to obtain the true gradient. In addition, the bias of the approximate centralized value function is large at the beginning of the training. Therefore, in practice, we can usually only obtain the sampled mean gradient with deviations. These deviations may cause the policy of agent $i$ to be updated in a sub-optimal direction, preventing the exploration of trajectories with higher returns. We call this problem: \textbf{The Policies Overfitting in Multi-agent Cooperation(POMAC)}. 

\begin{figure}[h]
\begin{subfigure}{.5\textwidth}
  \centering
  \includegraphics[height=2.8cm]{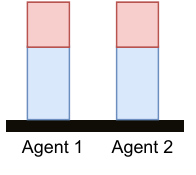}
  \caption{The stochastic policy gradient with the sampled shared advantage values.}
  \label{fig:gradient1}
\end{subfigure}
\begin{subfigure}{.5\textwidth}
  \centering
  \includegraphics[height=2.8cm]{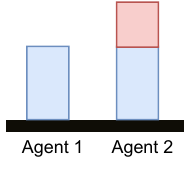}
  \caption{The true policy gradient.}
  \label{fig:gradient2}
\end{subfigure}
\caption{These bar charts represents the action probability.} {the red area indicates the amount of improvement of the probability by the policy gradient.}
\label{fig:gradient}
\end{figure}

To explain this problem more straightforwardly, we consider a multi-agent cooperative task with two agents. We assume that there is only one sample with reward $r_t$ and state $s_t$ in a batch, and the advantage value 
\begin{equation}
A^{\pi}\left(s_{t}, \vec{a_{t}}\right):= r_t + \gamma V^{\pi}\left(s_{t+1}\right) -V^{\pi}\left(s_{t}\right)
\end{equation}
is obtained by agent two and is not related to agent one \footnote{For example, agent one and agent two are far apart and agent two gets a reward $r_t$.}. The stochastic policy gradients with the this shared advantage value, i.e. $\frac{\partial \hat{J}}{\partial  \pi^i\left(a_{t}^i \mid s_{t}\right)} \propto A^{\pi}\left(s_{t}, \vec{a_{t}}\right)$, may improve probabilities of the policies of both agents, shown in Figure \ref{fig:gradient1}. But intuitively, the agent one's policy should not be updated as the advantage value is not related to agent one, shown in Figure \ref{fig:gradient2}.  By contrast, assuming we have an infinite number of samples to compute an unbiased advantage value: since agent one is independent of the advantage values under state $s_t$, the value of marginal advantage function (Eq. \ref{eqn:marginal}) of agent one will be equal to zero.

\begin{figure*}[htbp]
  \centering
  \includegraphics[width=1.0\textwidth]{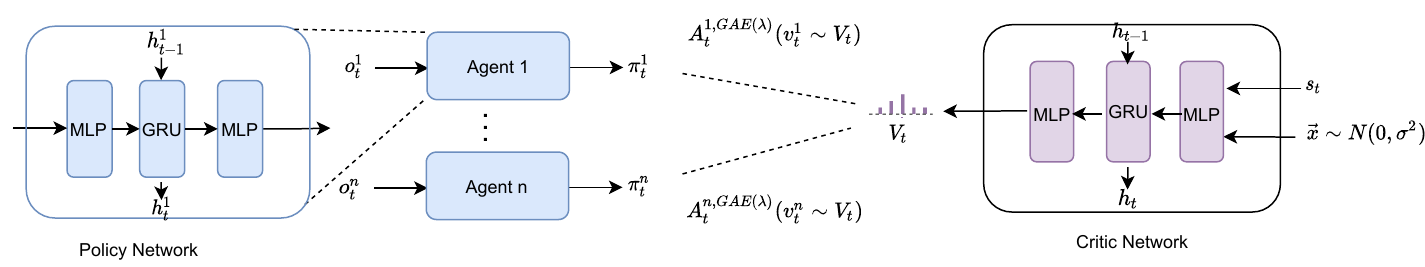}
  \caption{\centering{The framework of Noisy Value MAPPO (NV-MAPPO).}} {The noise of $V_t$ propagate to the advantage values $A_t$.}
  \label{fig:framework}
\end{figure*}

\section{Method} \label{sec:method}
\subsection{Noisy Advantage Values}
Intuitively, POMAC may be solved by explicit credit assignment; that is we can decompose the centralized advantage value to each agent $i$, i.e., $A(s, \vec{a}) = \sum_{i}^{N} A^i(s, a^i)$. In this way, the shared advantage value will do not affect unrelated agents. However, it is usually not easy to decompose the correct $A^i$ in a multi-agent system. Since the sampled advantage values are usually biased, our core motivation is to smooth these advantage values with noise to prevent multi-agnet policies overfitting caused by POMAC and environmental non-stationary, \textbf{likewise the label smoothing in image classification}. In this section, we propose two noisy advantage values methods for policy regularization,

\textbf{(I) Noisy-Advantage MAPPO (NA-MAPPO)} demonstrated in Algo. \ref{algo:adv-mappo} (Appendix \ref{appendix:mappo}). We sample a gaussian noise for each agent $i$,

\begin{equation}
    x^i \sim \mathcal{N}(0, 1), \forall i \in N
\end{equation}

where $N$ is the number of agents. Next, we mix the advantage values $A^b$ with the noises by a weight $\alpha$ (Eq. \ref{eqn:adv_mappo1}), perturbing the advantage values. Then, we can use these noisy advantage values to train multi-agent policies. 

\begin{equation}
A^i_b = (1- \alpha) \cdot A_b + \alpha \cdot x^i, \forall i \in N, b \in B
\label{eqn:adv_mappo1}
\end{equation}

\textbf{(II) Noisy-Value MAPPO (NV-MAPPO)} We randomly sample a gaussian noise vector $\vec{x}^i \sim \mathcal{N}(0, \sigma^2)$ for each agent $i$, where $\sigma^2$ is the variance can be seen as the noise intensity (we fine-tune $\sigma$ for each scenario, shown in Appendix \ref{sec:hyper}).Then we concatenate the noise $\vec{x}^i$ with global state $s$. As shown in Figure \ref{fig:framework}, we feed the concatenated features to the centralized value network to generate noise value $v^i$ for each agent $i$, 

\begin{equation}
v^i = V(concat(s, \vec{x}^i)), \forall i \in N
\label{eqn:adv_mappo}
\end{equation}

The random noise $\vec{x^i}$ disturbs the centralized value network and propagate to the advantage value $A^i = r + \gamma v^i(s_t) - v^i(s_{t+1})$, perturbing the advantage values. These advantage noises bring the following benefits,

\begin{enumerate}
    \item the advantage noises prevent the multi-agent policies over-fitting caused by the sampled advantage values with deviations and environmental non-stationarity.
    \item the policies trained by $N$ noisy value networks are similar to policies ensemble.
    \item the different noises $x^i$ of each agent drive the multi-agent policies go in different directions, which encourage agents to explore diverse trajectories.
\end{enumerate}

We then \textbf{combine the noisy value function with MAPPO, MAPG, and IPPO} to propose NV-MAPPO, Noisy-Value MAPG (NV-MAPG), and Noisy-Value IPPO (NV-IPPO), demonstrated in Algo. \ref{algo:nmappo} (Appendix \ref{appendix:mappo}). At last, we also show the difference between our method and MAPPO-FP in Table \ref{table:diff}.

\begin{table}[h]
\begin{tabular}{lcc}
\hline
Algo.                  & MAPPO-FP & Noisy-MAPPO \\ \hline
Expert agent-specific features     & Yes      & No          \\
Centralized Value-function & No       & Yes         \\
Noise                 & No       & Yes         \\ \hline
\end{tabular}
\caption{The difference between MAPPO-FP and Noisy-MAPPO.}
\label{table:diff}
\end{table}

\begin{table*}[htbp] 
\centering
\begin{tabular}{lcccccccc}
\hline
Senarios       & Difficulty     & NV-MAPPO        & NA-MAPPO       & NV-IPPO        & MAPPO  & MAPPO-FP       & IPPO           & Fine-tuned QMIX \\ \hline
2s3z           & Easy           & \textbf{100\%}  & -              & -              & \textbf{100\%} & \textbf{100\%} & \textbf{100\%} & \textbf{100\%}  \\
1c3s5z         & Easy           & \textbf{100\%}  & -              & -              & \textbf{100\%} & \textbf{100\%} & \textbf{100\%} & \textbf{100\%}  \\
3s5z           & Easy           & \textbf{100\%}  & -              & -              & \textbf{100\%} & \textbf{100\%} & \textbf{100\%} & \textbf{100\%}  \\
2s\_vs\_1sc    & Easy           & \textbf{100\%}  & -              & -              & \textbf{100\%} & \textbf{100\%} & \textbf{100\%} & \textbf{100\%}  \\
3s\_vs\_5z     & Hard           & \textbf{100\%}  & \textbf{100\%} & \textbf{100\%} & 98\%           & \textbf{100\%} & \textbf{100\%} & \textbf{100\%}  \\
2c\_vs\_64zg   & Hard           & \textbf{100\%}  & \textbf{100\%} & \textbf{100\%} & \textbf{100\%} & \textbf{100\%} & 98\%           & \textbf{100\%}  \\
5m\_vs\_6m     & Hard           & 89\%            & 85\%           & 87\%           & 25\%           & 89\%           & 87\%           & \textbf{90\%}   \\
8m\_vs\_9m     & Hard           & 96\%            & 96\%           & 96\%           & 93\%           & 96\%           & 96\%           & \textbf{100\%}  \\
MMM2           & Super Hard     & 96\%            & 96\%           & 86\%           & 96\%           & 90\%           & 86\%           & \textbf{100\%}  \\
3s5z\_vs\_3s6z & Super Hard     & 87\%            & 72\%           & \textbf{96\%}  & 56\%           & 84\%           & 82\%           & 75\%(env=8)     \\
6h\_vs\_8z     & Super Hard     & 91\%            & 90\%           & \textbf{94\%}  & 15\%           & 88\%           & 84\%           & 91\%            \\
corridor       & Super Hard     & \textbf{100\%}  & \textbf{100\%} & 98\%           & 3\%            & \textbf{100\%} & 98\%           & \textbf{100\%}  \\
27m\_vs\_30m   & Super Hard     & \textbf{100\%}  & 98\%           & 72\%           & 98\%           & 94\%           & 69\%           & \textbf{100\%}  \\
Avg. Score     & \textbf{Hard+} & \textbf{95.5\%} & 93.2\%         & 91.9 \%        & 64.9\%         & 93.4\%         & 88.8\%         & 95.1\%          \\ \hline
\end{tabular}
\caption{Median test win percentage of MARL algorithms in all scenarios.} {The test results for MAPPO-FP and IPPO are from \cite{yu2021surprising}.}
\label{table:benchmarks}
\end{table*}

\section{Experinments} \label{sec:experi}
In this section, we first evaluate the performance of NV-MAPPO, NA-MAPPO, and NV-IPPO in SMAC; and we analyze how these noises affect their performance and the entropy of the policies of MAPPO. We then evaluate the expressive power of NV-MAPPO on two non-monotonic matrix games.

\subsection{Benchmark Environments}
\subsubsection{Starcraft Multi-agent Challenge (SMAC)} \citep{samvelyan2019starcraft} focuses on micromanagement challenges where each unit is controlled by an independent agent that must act based on local observations, which has become a common-used benchmark for evaluating state-of-the-art MARL approaches, such as \citep{rashid2018qmix, son2019qtran, mahajan2020maven, foerster2017counterfactual}. SMAC offers diverse sets of scenarios, which are classified as Easy, Hard, and Super Hard scenarios. We use the hardest scenarios in SMAC as our main benchmark environment.

\subsubsection{Non-monotonic Matrix Game}  \citep{son2019qtran} \citep{mahajan2020maven} show the non-monotonic matrix games that violates the monotonicity constraint. For the matrix game Table \ref{tb:nonmatrixa} (Sec. \ref{sec:nonmono}); in order to obtain the reward 8, both agents must select the first action 0 (actions are indexed from top to bottom, left to right); if only one agent selects action 0, they obtain reward -12. QMIX learns incorrect $Q_{tot}$ in such non-monotonic matrix games \citep{son2019qtran} \citep{mahajan2020maven}. We use two payoff matrices (Sec. \ref{sec:nonmono}, Table \ref{tb:nonmatrixa} and \ref{tb:nonmatrixb}) to evaluate the expressive power of NV-MAPPO.

\subsubsection{Evaluation Metric}
Our primary evaluation metric is the function that maps the steps for the environment observed throughout the training to the median test-winning percentage/median test return of the evaluation. Just as in QMIX~\citep{rashid2018qmix}, we repeat each experiment with several independent training runs (five independent random experiments). 

\subsection{SMAC}
In this section, we evaluate the performance of the algorithms on SMAC. We test our noisy value function on MAPPO and IPPO, i.e., NV-MAPPO nad NV-IPPO, respectively, in SMAC. We use the Fine-tuned QMIX \citep{hu2021riit} and MAPPO-FP as the baseline, as they achieve SOTA performance in SMAC among the previous works; we do not compare NV-MAPPO with MADDPG as the past experiments \citep{peng2020facmac,zhou2020learning} shows that it does not perform well under SMAC.

\subsubsection{Performance Comparison}
The experimental results in Table \ref{table:benchmarks} demonstrate that (1) \textbf{performance of NV-MAPPO significantly exceeds that of MAPPO} on most hard scenarios \footnote{\citep{samvelyan2019starcraft} illustrates that some of these hard scenarios are more difficult to explore.}, such as 5m\_vs\_6m (+65\%), corridor (+97\%), 6h\_vs\_8z (+87\%) and 3s5z\_vs\_3s6z (+31\%). (2) NV-IPPO achieves extraordinarily high win rates in Super Hard scenarios 3s5z\_vs\_3s6z (96\%) and 6h\_vs\_8z (94\%); we speculate that this is because the noise also prevents IPPO from overfitting due to non-stationarity. (3) The average performance of NV-MAPPO on hard scenarios is better than that of Fine-tuned QMIX and MAPPO-FP. (4) We compare MAPG and NV-MAPG in the Appendix \ref{appendix:all_result} and show that NV-MAPG also performs significantly better than MAPG.

All these results indicate that the noisy value function works well in practical tasks.
Since we use Fine-tuned QMIX \citep{hu2021riit} as the baseline, the median test-winning rates of QMIX are significantly better than the experimental results in the past literature \citep{samvelyan2019starcraft,rashid2018qmix,yang2020qatten,mahajan2020maven}. So far, \textbf{NV-MAPPO and NV-IPPO together achieve SOTA in SMAC}. Specifically, NV-IPPO (for 3s5z\_vs\_3s6z and 6h\_vs\_8z) and NV-MAPPO (for other hard scenarios) have an \textbf{average win rate of 97\% for all hard scenarios}.

\begin{figure}[htbp]
  \centering
  \includegraphics[width=0.9\columnwidth]{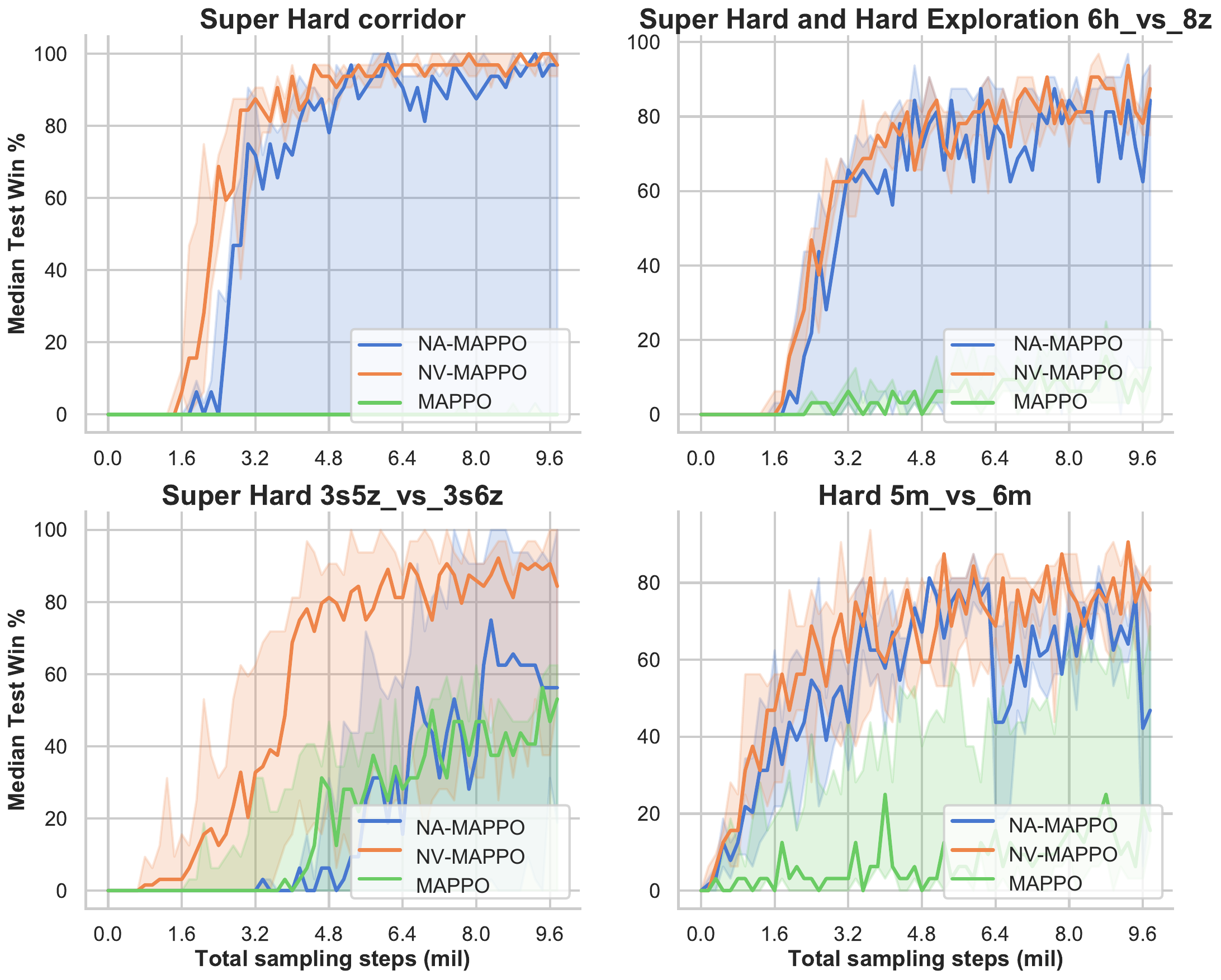}
  \caption{\centering{Comparing Noisy-Advantage MAPPO (NA-MAPPO) with Noisy-Value MAPPO (NV-MAPPO).}} {The win rates of Noisy-Advantage MAPPO have a large variance.}
  \label{fig:adv_noist}
\end{figure}
\subsubsection{Comparing NA-MAPPO with NV-MAPPO} \label{sec:comparing_noisy} In previous sections, We have proposed two noise-based methods, i.e, NA-MAPPO with NV-MAPPO, to resolve the POMAC. In this section, we compare their performance in SMAC. As shown in Figure \ref{fig:adv_noist}, we find that the Noisy-Advantage method may harm the stability of the algorithm in some scenarios, i.e, \textbf{the win rates of Noisy-Advantage methods have a large variance}. We speculate that it may be the explicit noises destroy the original direction of the policy gradient. However, the performance of the NA-MAPPO is still comparable to NV-MAPPO in some hard scenarios of SMAC; and we note that the NA-MAPPO is extremely easy to implement. All of these results indicate the noise advantage values do improve the performance of vanilla MAPPO.

\begin{figure}[htbp]
  \centering
  \includegraphics[width=0.9\columnwidth]{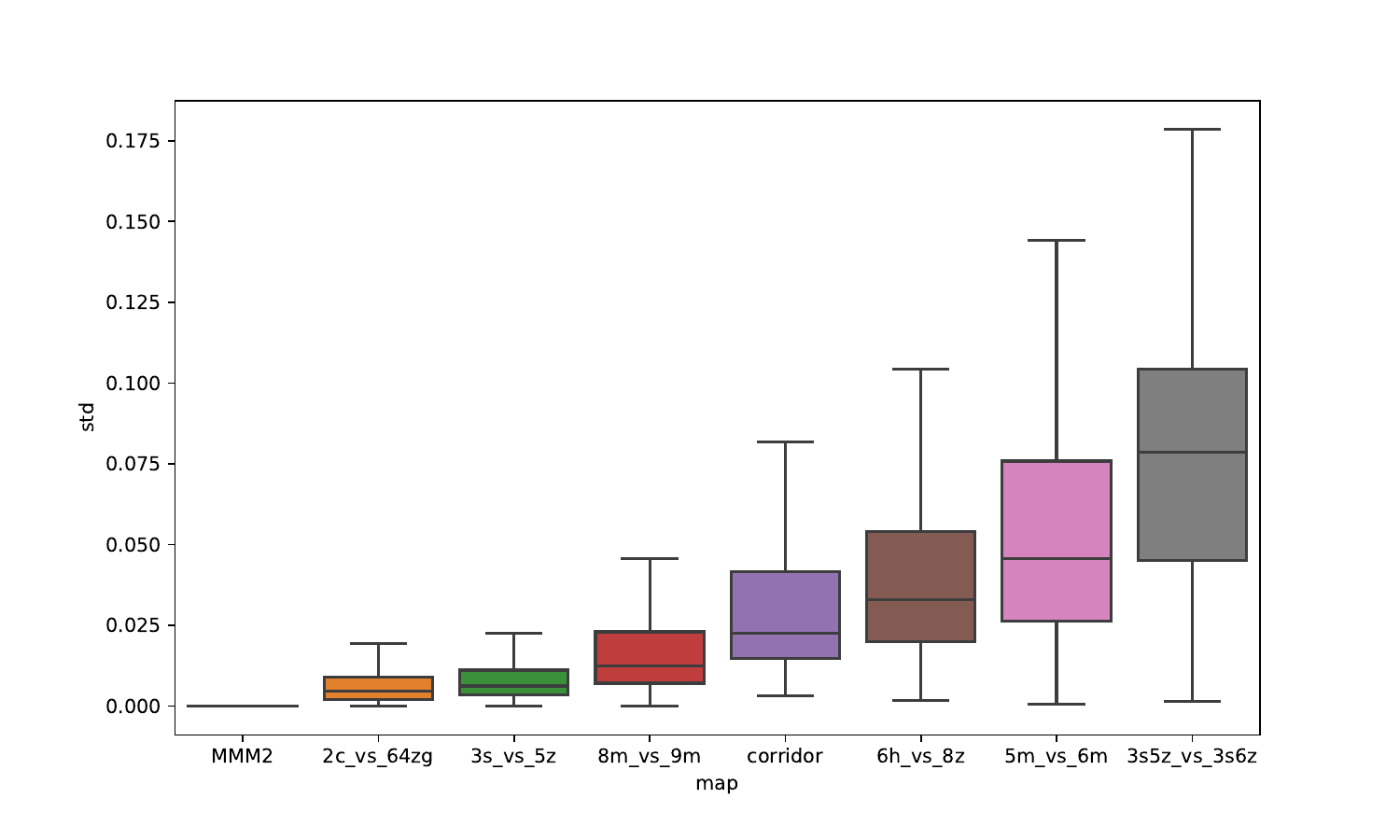}
  \caption{\centering{The standard deviation of noisy value function $v^i$ in the agent dimension.}} {The figure shows that scenarios with large variance of $v^i$ imply that noise also has a significant performance improvement on them.}
  \label{fig:variance}
\end{figure}
\subsubsection{Analysis of the Variance of $v^i$}  Next, we perform further experimental analysis on how the noisy value function of NV-MAPPO affects the performance. We show the standard deviation of the value function $v^i$ in agent dimension for some Hard scenarios in Figure \ref{fig:variance}. We find that \textbf{the large variance of $v^i$ in some scenarios implies that the performance improvement of NV-MAPPO over vanilla MAPPO in these scenarios is also large}, such as 3s5z\_vs\_3s6z and 6h\_vs\_8z (see Figure \ref{fig:variance} and Figure \ref{fig:adv_noist}). This law reveals that the performance improvement of NV-MAPPO does come from noise perturbation of value function. 

\begin{figure}[htbp]
  \centering
  \includegraphics[width=0.9\columnwidth]{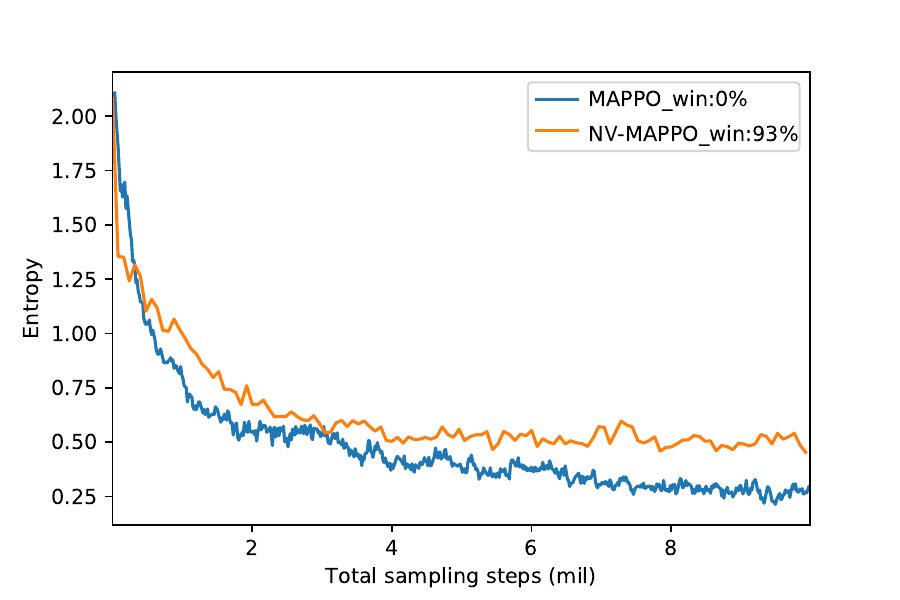}
  \caption{\centering{The average entropy of the polices of vanilla MAPPO and NV-MAPPO.}}
  \label{fig:entropy}
\end{figure}
\subsubsection{Analysis of Policy Entropy} Finally we analyze the effect of the noisy value function on the entropy of the policies on scenario $3s5z\_vs\_3s6z$. As shown in Figure \ref{fig:entropy}, the entropy of vanilla MAPPO's policies drops rapidly and falls into the local optimal solution, thus the winning rate is always zero. As for NV-MAPPO, we smoothed the sampled advantage values and the noise prevents policies overfitting, thus the entropy of the policies decreases more cautiously.

\subsection{Non-monotonic Matrix Game} \label{sec:nonmono} 
\begin{figure}[htbp]
\begin{minipage}[t]{0.46\columnwidth}
  \centering
  \includegraphics[width=1.0\columnwidth]{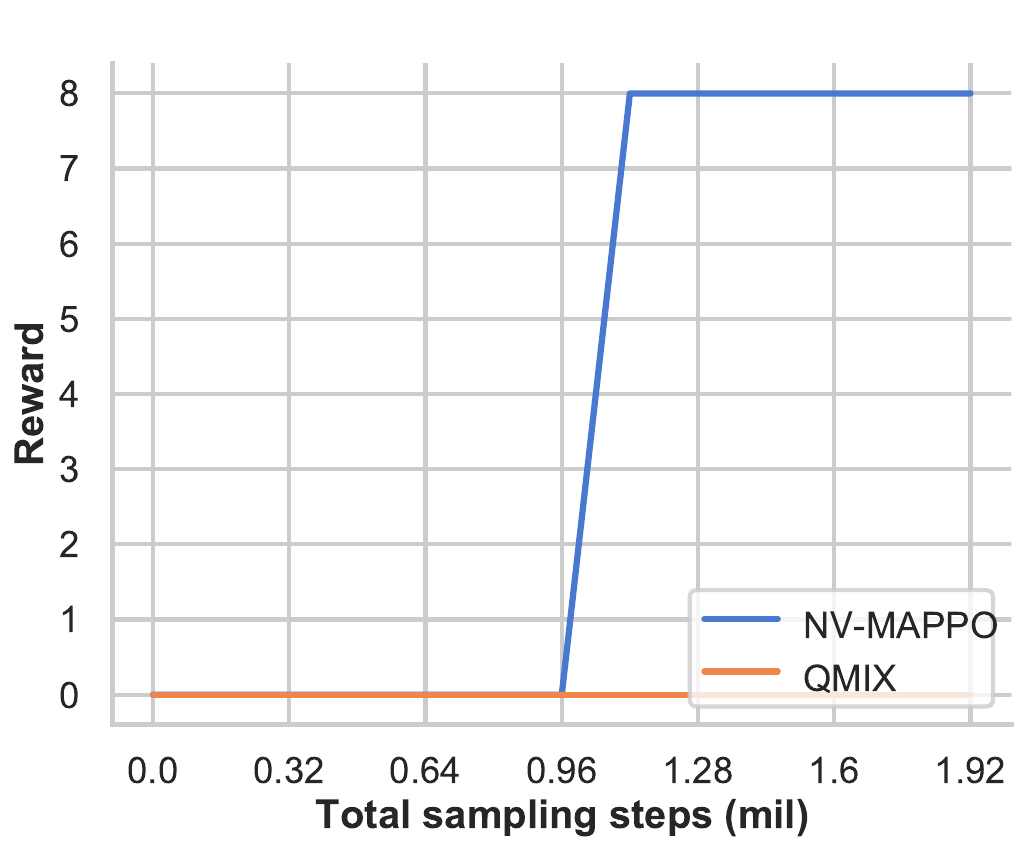}
  \subcaption[]{\centering{Non-monotonic matrix game, Table \ref{tb:nonmatrixa}}}
  \label{fig:nonmonotonica}
\end{minipage}
\begin{minipage}[t]{0.46\columnwidth}
  \centering
  \includegraphics[width=1.0\columnwidth]{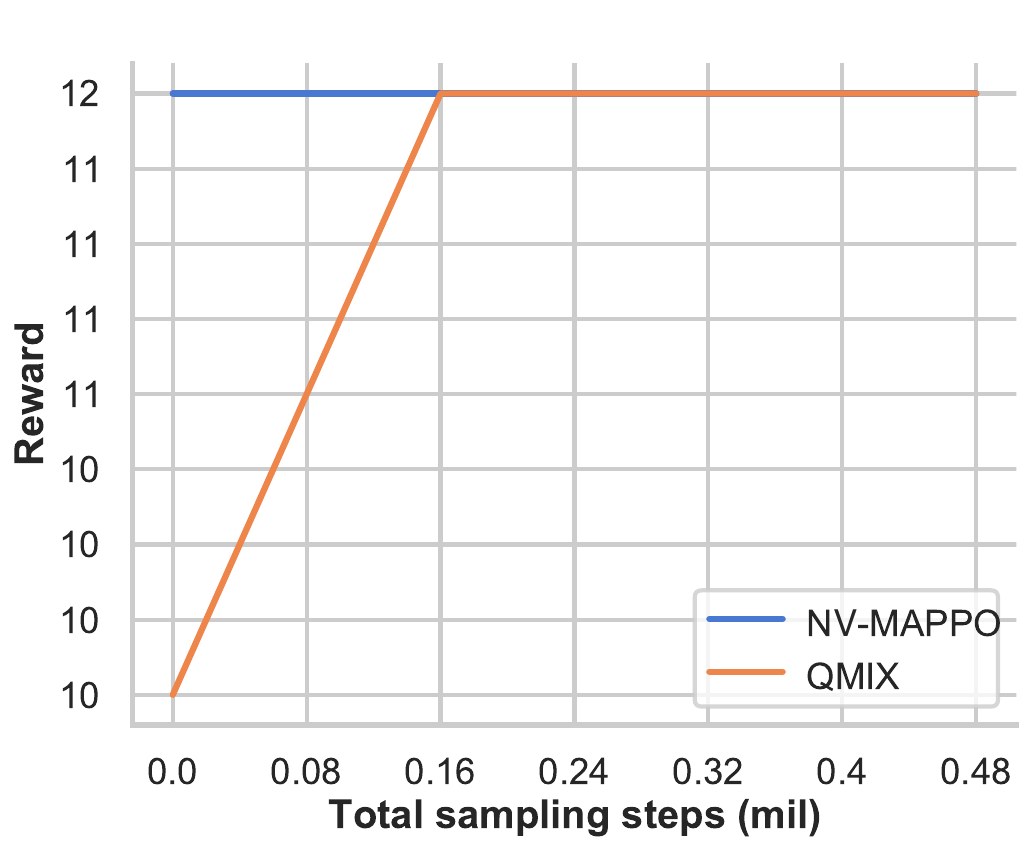}
  \subcaption[]{\centering{Non-monotonic matrix game, Table \ref{tb:nonmatrixb}}}
  \label{fig:nonmonotonicb}
\end{minipage}
\caption{\centering{Test returns for non-monotonic matrix games from Sec.\ref{sec:nonmono}}}
\end{figure}

In this section, we evaluate the expressiveness of NV-MAPPO using two non-monotonic matrix games; As shown in Figure \ref{fig:nonmonotonica} and \ref{fig:nonmonotonicb}, since there are no constraints on the value function of MAPPO (e.g., monotonicity constraints), the test performance of NV-MAPPO in both of these non-monotonic games are significantly better than QMIX. Since we use the Fine-tuned QMIX, the test returns of QMIX in matrix \ref{tb:nonmatrixb} is better than that in the past literature \citep{mahajan2020maven}.

\begin{table}[htbp]
 \begin{minipage}{0.48\columnwidth}
  \centering
      \begin{tabular}{|p{.5cm}|p{.5cm}|p{.5cm}|}
        \hline \ 8 & -12 & -12 \\
        \hline -12 & 0 & 0 \\
        \hline -12 & 0 & 0 \\
        \hline
    \end{tabular}
      \subcaption{Payoff matrix 1}
      \label{tb:nonmatrixa}
  \end{minipage}
  \begin{minipage}{0.48\columnwidth}
  \centering
         \begin{tabular}{|p{.5cm}|p{.5cm}|p{.5cm}|}
        \hline 12 & 0 & 10 \\
        \hline 0 & 10 & 10 \\
        \hline 10 & 10 & 10 \\
        \hline
        \end{tabular}
        \subcaption{Payoff matrix 2}
        \label{tb:nonmatrixb}
  \end{minipage}
  \caption{Non-monotonic matrix games from \cite{son2019qtran} (a) and \cite{mahajan2020maven}(b)}
\end{table}

\section{Conclusion} In this paper, we propose two noisy advantage-value methods (NV-MAPPO and NA-MAPPO) to mitigate the overfitting of multi-agent policies. The experimental results show that NV-MAPPO and NV-IPPO together achieve extraordinarily high win rates in all scenarios and achieve SOTA in SMAC, without limitation of expressiveness and artificial agent-specific features. Our work demonstrates that the perturbation of policies with noisy advantage values effectively improves the performance of the multi-agent actor-critic algorithms in some scenarios.


\clearpage
\bibliographystyle{ACM-Reference-Format} 
\bibliography{main}
\clearpage
\appendix

\begin{figure*}[thbp]
  \centering
  \includegraphics[width=0.9\textwidth]{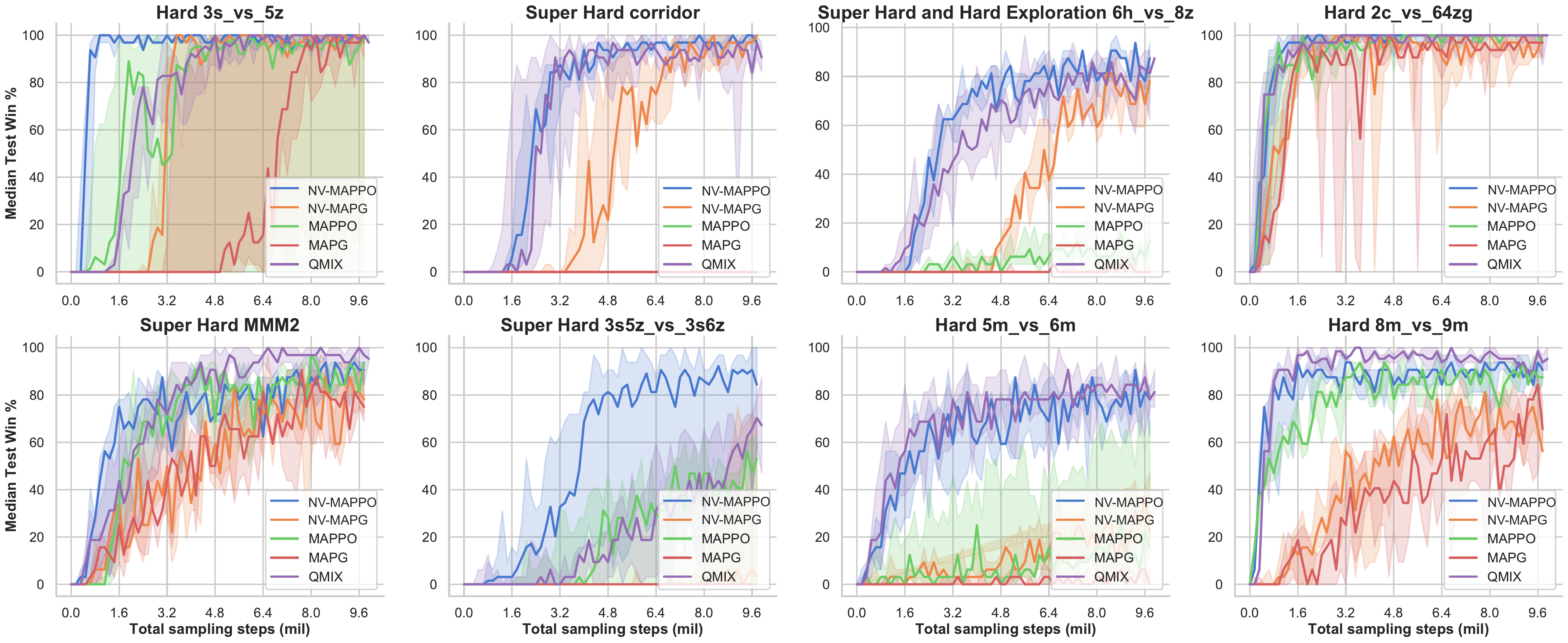}
  \caption{\centering{Median test win rate of MARL algorithms on hard scenarios in SMAC.}} {NV denotes Noisy-Value}
  \label{fig:baselines}
\end{figure*}
\begin{figure*}[htbp]
  \centering
  \includegraphics[width=0.75\textwidth]{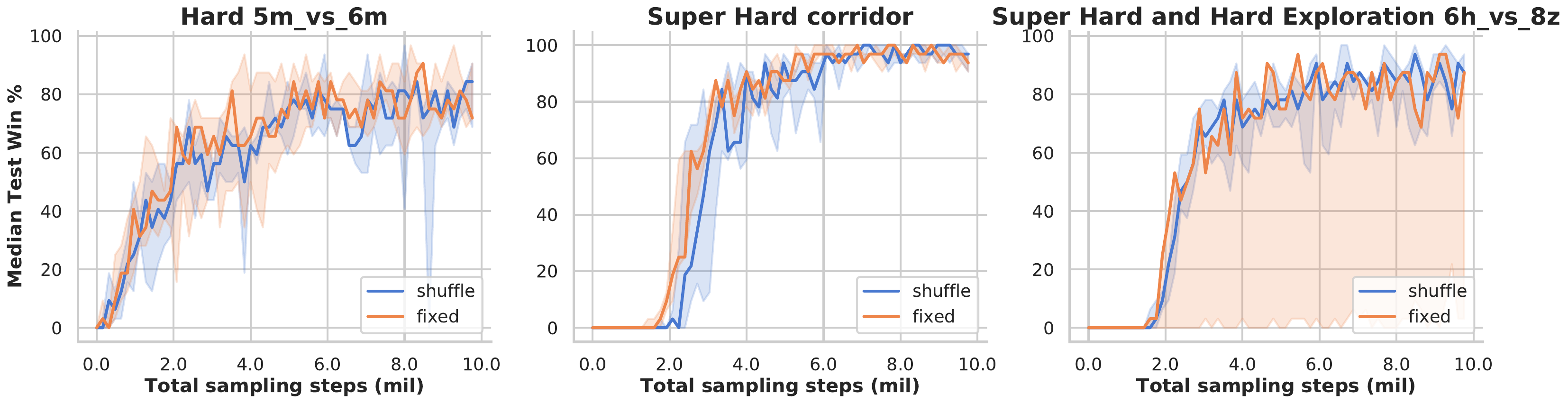}
  \caption{\centering{The performance comparison of fixed noise vectors and shuffle noise vectors (every 100 episodes).}}
  \label{fig:shuffle}
\end{figure*}

\section{Pseudocode} \label{appendix:mappo} 
Algo. \ref{algo:nmappo} and Algo. \ref{algo:adv-mappo} demonstrate NV-MAPPO and NA-MAPPO, respectively. NV-MAPPO adds Gaussian noise to the input layer of the Value Network, and NA-MAPPO adds Gaussian noise directly to the normalized advantage values.

\section{Experinmental Details}

\subsection{Ommited Figures}
\label{appendix:all_result} Here, we echo the experiments in Sec. \ref{sec:experi}. Figure \ref{fig:baselines} shows the experimental results for Fine-tuned QMIX, MAPPO, MAPG, NV-MAPPO, and NV-MAPG, which demonstrates the performance gains from noise (MAPG vs NV-MAPG, MAPPO vs NV-MAPPO). 

\subsection{Hyperparameters} \label{sec:hyper}
Our hyperparameters are hevily based on recent papers \cite{yu2021surprising} and \cite{hu2021riit}, who fine-tune PPO \footnote{PPO Code: \url{https://github.com/marlbenchmark/on-policy}} and QMIX \footnote{QMIX code: \url{https://github.com/hijkzzz/pymarl2}}, respectively, to make them work well in complex multi-agent tasks, such as SMAC. Table \ref{table:comm_hyper} shows the common hyperparameters of QMIX and MAPPO. Table \ref{table:hyper} shows the hyperparameters of NV-MAPPO and NA-MAPPO for each scenarios, where the values of $\sigma$ and $\alpha$ are depend on the scenarios.  

For the \textbf{noise shuffle interval} of NV-MAPPO, we find that the performance of the fixed gaussian noise vectors $\vec{x^i}$ is comparable to that of the noise of periodic shuffle (100 episodes), as shown in Figure \ref{fig:shuffle}. In addition, \textbf{we do not recommend frequent noise updates} because it may destroy the stability of the learning of algorithms. The idea of periodic noise updates comes from the periodic target network updates in Deep Q-networks (DQN) \cite{mnih2013playing}. Note that the fixed noise vector cannot be seen as an identifier for an agent, as the value network cannot infer which agent it is just by a noise vector $\vec{x}$ and state $s$ (unless you also feeds the observation $o^i$ of the agent, likewise MAPPO-FP).

\textbf{Other settings} For the non-monotonic matrix games, we set the number of environments of all algorithms to $32$, buffer length to $1$, noise vector dim to $10$, training epochs to $10$, and $\sigma$ to 1. At last, we use StarCraft 2 (SC2.4.10) in the latest PyMARL in our experiments.

\makeatletter
\newcommand{\removelatexerror}{\let\@latex@error\@gobble}
\makeatother

\newcommand{\myalgorithma}{%
\begingroup
\removelatexerror

\begin{algorithm*}[H]
  \SetKwData{Left}{left}\SetKwData{This}{this}\SetKwData{Up}{up}
  \SetKwInOut{Input}{input}\SetKwInOut{Output}{output}

  \Input{Initialize parameters $\theta; \phi$;
  $\mathcal{D}\leftarrow \{\}$;
  batch size $B$; $N$ agents;
  noise variance $\sigma^2$;\\ entropy loss weight $\eta$; $\lambda$ for GAE($\lambda$)\;}
  \BlankLine
    Sample random noise vectors $\vec{x}^i$ for each agent, $\vec{x}^i\sim \mathcal{N}(0,\sigma^2),  \forall i \in N$\; 
  \For{each episodic iteration}{
    \For{episodic step t}{\label{forins}
        $\vec{a_t} = [\pi^i_\theta(o_t^i), \forall i \in N$]\;
        Execute actions $\vec{a_t}$, observe $r_t,s_{t+1},o_{t+1}$\;
        $\mathcal{D}\leftarrow\mathcal{D}\cup\mathcal{D}\{(s_t,\vec{o_t},\vec{a_t},r_t,s_{t+1},\vec{o_{t+1},}\}$\;
    }
    \If{at noise vectors shuffle interval} {
        Shuffle the noise vectors $\vec{x}^i$ in agent dimension.
    }
    
    Sample random batch $B$ from $\mathcal{D}$\;
    Noise value function forward for each agent, $v^i_b(\phi) = V_\phi(concat(s_b, \vec{x}^i)), \forall i \in N, b \in B$\;
    
    Compute advantage $\hat{A_1^i},\ldots,\hat{A_b^i}$ and returns $\hat{R_1^i},\ldots,\hat{R_b^i}$ via GAE($\lambda$) with $v^i_b(\phi), \forall i \in N, b \in B$\;
    
  \For{each training epochs}{
    Update critic by minimizing the loss $L(\phi)$\;
    \begin{equation}
        L(\phi)= \frac{1}{B \cdot N}\sum_{i=1}^{B}\sum_{i=1}^N(v^i_b(\phi)-\hat{R}^i_b)^2 \nonumber
    \end{equation}
    
    Update policy by using loss $L(\theta)$\;
    \begin{eqnarray}
    \begin{aligned}
        r_{b}^{i}(\theta) &=\frac{\pi^i_\theta(a_b^i|o_b^i)}{\pi^i_{\theta_{old}}(a_b^i|o_b^i)}, \forall i \in N, b \in B \nonumber \\
        L(\theta)= & \frac{1}{B\cdot N}\sum_{b=1}^{B}\sum_{i=1}^N [ \min(r_{b}^{i}(\theta)\hat{A}_b^{i},clip(r_{b}^{i}(\theta),1-\epsilon,1+\epsilon)\hat{A}_b^{i}) \nonumber \\
         &-\eta \mathcal{H}(\pi^i_\theta(o_b^{i})) ] \nonumber
    \end{aligned}
    \end{eqnarray}
    where $\mathcal{H}$ is the Shannon Entropy.
    }
  }
  \caption{NV-MAPPO}\label{algo:nmappo}
\end{algorithm*}
\endgroup}

\newcommand{\myalgorithmb}{%
\begingroup
\removelatexerror%
\begin{algorithm*}[H]
  \SetKwData{Left}{left}\SetKwData{This}{this}\SetKwData{Up}{up}
  \SetKwInOut{Input}{input}\SetKwInOut{Output}{output}

  \Input{Initialize parameters $\theta; \phi$;
  $\mathcal{D}\leftarrow \{\}$;
  batch size $B$; $N$ agents;
  noise weight $\alpha$;\\ entropy loss weight $\eta$; $\lambda$ for GAE($\lambda$)\;}
  \BlankLine
  Sample Gaussian noise $x^i \sim \mathcal{N}(0, 1), \forall i \in N$;\\
  \For{each episodic iteration}{
    \For{episodic step t}{\label{forins}
        $\vec{a_t} = [\pi^i_\theta(o_t^i), \forall i \in N$]\;
        Execute actions $\vec{a_t}$, observe $r_t,s_{t+1},o_{t+1}$\;
        $\mathcal{D}\leftarrow\mathcal{D}\cup\mathcal{D}\{(s_t,\vec{o_t},\vec{a_t},r_t,s_{t+1},\vec{o_{t+1},}\}$\;
    }
    Sample random batch $B$ from $\mathcal{D}$\;
    Compute advantage $\hat{A_1},\ldots,\hat{A_b}$ and returns $\hat{R_1},\ldots,\hat{R_b}$ via GAE($\lambda$);\\
    then mixing the noise with the normalized advantage values:\\
    \begin{equation}
        \hat{A^i_b} = (1- \alpha) \hat{A_b} + \alpha \cdot x^i, \forall i \in N, b \in B \nonumber
    \end{equation}
    
  \For{each training epochs}{
    Update critic by minimizing the loss $L(\phi)$\;
    \begin{equation}
        L(\phi)= \frac{1}{B}\sum_{b=1}^{B}(v_b(\phi)-\hat{R}_b)^2 \nonumber
    \end{equation}
    
    Update policy by using loss $L(\theta)$\;
    \begin{eqnarray}
    \begin{aligned}
        r_{b}^{i}(\theta) &=\frac{\pi^i_\theta(a_b^i|o_b^i)}{\pi^i_{\theta_{old}}(a_b^i|o_b^i)}, \forall i \in N, b \in B \nonumber \\
        L(\theta)= & \frac{1}{B\cdot N}\sum_{b=1}^{B}\sum_{i=1}^N [ \min(r_{b}^{i}(\theta)\hat{A}_b^{i},clip(r_{b}^{i}(\theta),1-\epsilon,1+\epsilon)\hat{A}_b^{i}) \nonumber \\
         &-\eta \mathcal{H}(\pi^i_\theta(o_b^{i})) ] \nonumber
    \end{aligned}
    \end{eqnarray}
    where $\mathcal{H}$ is the Shannon Entropy.\\
    }
  }
  \caption{NA-MAPPO}\label{algo:adv-mappo}
\end{algorithm*}
\endgroup
}

\begin{figure*}[htbp]
\begin{minipage}[t]{0.48\textwidth}
\myalgorithmb
\end{minipage}
\quad
\begin{minipage}[t]{0.48\textwidth}
\myalgorithma
\end{minipage}
\end{figure*}

\begin{table*}[htbp] 
\centering
\begin{tabular}{lcc}
\hline
hyperparameters      & MAPPO and MAPG    &QMIX\\ \hline
num envs             & 8            & 8\\          
buffer length        & 400          &-\\
batch size(episodes) & -            &128\\
num GRU layers       & 1            &1\\
RNN hidden state dim & 64           &64\\
fc layer dim         & 64           &64\\
num fc before RNN               & 1            &1\\
num fc after RNN    & 1            &1\\
num noise dim        & 10           &-\\
Adam \cite{kingma2014adam} lr                   & 5e-4         &1e-3\\
Q($\lambda$)        & -         &0.6, (0.3 for 6h\_vs\_8z) \\
GAE($\lambda$)       & 0.95       & -\\
entropy coef       & 0.01       & -\\
PPO clip       & 0.2       & -\\
noise shuffle interval (episodes) & $+\infty$ & -\\
$\epsilon$ anneal steps       & -       & 100k, (500k for 6h\_vs\_8z)\\
\hline
\end{tabular}
\caption{Common hyperparameters used in the SMAC domain for all algorithms.}
\label{table:comm_hyper}
\end{table*}
\begin{table*}[htbp]
\centering
\begin{tabular}{@{}llccccccccc@{}}
\toprule
\multirow{2}{*}{map} & \multirow{2}{*}{PPO epochs} & \multirow{2}{*}{mini-batch} & \multirow{2}{*}{gain} & \multirow{2}{*}{network} & \multirow{2}{*}{stacked frames} & \multicolumn{1}{l}{NV-MAPPO} & \multicolumn{1}{l}{NV-MAPG} & NV-IPPO  & NA-MAPPO &  \\ \cmidrule(l){7-11} 
                     &                             &                             &                       &                          &                                 & $\sigma$                     & $\sigma$                    & $\sigma$ & $\alpha$ &  \\ \midrule
2s3z                 & 15                          & 1                           & 0.01                  & rnn                      & 1                               & 1                            & 1                           & -        & -        &  \\
1c3s5z               & 15                          & 1                           & 0.01                  & rnn                      & 1                               & 1                            & 1                           & -        & -        &  \\
3s5z                 & 5                           & 1                           & 0.01                  & rnn                      & 1                               & 1                            & 1                           & -        & -        &  \\
2s\_vs\_1sc          & 15                          & 1                           & 0.01                  & rnn                      & 1                               & 1                            & 1                           & -        & -        &  \\
3s\_vs\_5z           & 15                          & 1                           & 0.01                  & mlp                      & 4                               & 1                            & 1                           & 1        & 0.05     &  \\
2c\_vs\_64zg         & 5                           & 1                           & 0.01                  & rnn                      & 1                               & 1                            & 1                           & 1        & 0.05     &  \\
5m\_vs\_6m           & 10                           & 1                           & 0.01                  & rnn                      & 1                               & 8                           & 3                           & 0        & 0.05     &  \\
8m\_vs\_9m           & 15                          & 1                           & 0.01                  & rnn                      & 1                               & 1                            & 0.05                        & 1        & 0.05     &  \\
corridor             & 5                           & 1                           & 0.01                  & mlp                      & 1                               & 3                            & 1                           & 1        & 0.06     &  \\
MMM2                 & 5                           & 2                           & 1                     & rnn                      & 1                               & 0                            & 0.5                         & 0        & 0        &  \\
3s5z\_vs\_3s6z       & 5                           & 1                           & 0.01                  & rnn                      & 1                               & 10                           & 1                           & 8        & 0.05     &  \\
6h\_vs\_8z           & 5                           & 1                           & 0.01                  & mlp                      & 1                               & 1                            & 1                           & 1        & 0.06     &  \\
27m\_vs\_30m         & 5                           & 1                           & 0.01                  & rnn                      & 1                               & 1                            & 1                           & 1        & 0        &  \\ \bottomrule
\end{tabular}
\caption{Hyperparameters for NV-MAPPO, NA-MAPPO, NV-IPPO, NV-MAPG and vanilla MAPPO in SMAC.}
\label{table:hyper}
\end{table*}


\end{document}